\documentclass[12pt]{iopart}

\usepackage{graphicx}
\usepackage{dcolumn}
\usepackage{bm}

\begin{document}
\title{Electronic and Geometric Corrugation of Periodically Rippled, Self-nanostructured Graphene Epitaxially Grown on Ru(0001)}

\author{Bogdana Borca$^1$, Sara Barja$^{1,2}$, Manuela Garnica$^{1,2}$, Marina Minniti$^2$, Antonio Politano$^2$, Josefa M. Rodriguez-Garc\'{\i}a$^1$, Juan Jose Hinarejos$^{1,3}$, Daniel Far\'{\i}as$^{1,3}$, Amadeo L. V\'{a}zquez de Parga$^{1,2,3}$ and Rodolfo Miranda$^{1,2,3}$}

\address{$^{1}$Departamento de F\'{i}sica de la Materia Condensada, Universidad Aut\'onoma de Madrid, Cantoblanco 28049, Madrid, Spain}
\address{$^{2}$Instituto Madrile\~no de Estudios Avanzados en Nanociencia (IMDEA- Nanociencia), Cantoblanco 28049, Madrid, Spain}
\address{$^{3}$Instituto de Ciencia de Materiales "Nicol\'{a}s Cabrera", Universidad Aut\'{o}noma de Madrid. Cantoblanco 28049, Madrid, Spain}

\ead{al.vazquezdeparga@uam.es}

\date{\today}

\begin{abstract}
Graphene epitaxially grown on Ru(0001) displays a remarkably ordered pattern of hills and valleys in Scanning Tunneling Microscopy (STM) images. The extent to which the observed "ripples" are structural or electronic in origin have been much disputed recently. A combination of ultrahigh resolution STM images and Helium Atom diffraction data shows that i) the graphene lattice is rotated with respect to the lattice of Ru and ii) the structural corrugation as determined from He diffraction is substantially smaller (0.15 {\AA}) than predicted (1.5 {\AA}) or reported from X-Ray Diffraction or Low Energy Electron Diffraction. The electronic corrugation, on the contrary, is strong enough to invert the contrast between hills and valleys above +2.6 V as new, spatially localized electronic states enter the energy window of the STM. The large electronic corrugation results in a nanostructured periodic landscape of electron and holes pockets.
\end{abstract}



\maketitle

\section{Introduction}

The growth of large, highly perfect epitaxial graphene monolayers is a prerequisite for most practical applications of this promising material \cite{geim1,geim2}. In addition, it is crucial to understand the interaction of graphene with the surfaces of substrates of different nature (oxides, semiconductors or metals), as well as with adsorbed molecules, in view of the sensitivity of the conduction properties of graphene to them \cite{schedin1}. Nanostructuring graphene (in stripes, dots or by periodic potentials), in turn, may reveal new physical phenomena and fascinating applications \cite{park1}. Most of these topics can be characterized in detail in what has become a benchmark system for epitaxial graphene: a self-organized, millimeter-large, periodically "rippled" epitaxial monolayer of graphene grown by soft Chemical Vapor Deposition under Ultra High Vacuum (UHV) conditions on single crystal metal substrates with hexagonal symmetry, such as Ru(0001) \cite{marchini1,amadeo1,sutter1}, Ir(111) \cite{diaye1,coraux1} or Pt(111) \cite{fujita1}. The superb control that allows the UHV environment facilitates a precise characterization of the system down to the atomic scale.

The difference in lattice parameter between graphene (gr) and the different metal substrates causes the appearance of moir\'{e} patterns with a range of apparent vertical corrugations and lateral periodicities with respect to the basic graphene structure \cite{marchini1,amadeo1,sutter1,diaye1,coraux1,fujita1}. In-situ STM imaging of graphene monolayers on Ru(0001) reveals periodic corrugations with (12$\times$12)\cite{marchini1,amadeo1,sutter1} and surface X-ray diffraction (XRD) experiments give a (25$\times$25) periodicity based on the distortion of the first Ru layer under the graphene \cite{martoccia1}. The carbon atoms on the gr/Ru(0001) are electronically inequivalent as reflected in the X-Ray Photoelectron Spectroscopy C 1s core levels, which show two peaks, i.e. two differently bonded C atoms \cite{preobajenski}.

The apparent amplitude in STM of the corrugation of the ripples for gr/Ru(0001) decreases from 1.1 {\AA} to 0.5 {\AA} when the tunneling bias goes from -0.8 V to 0.8 V \cite{amadeo2}. While STM studies \cite{amadeo1} ascribed most of the apparent corrugation to electronic effects in a weakly ($<$0.3 {\AA}) structurally corrugated graphene overlayer, Density Functional Theory (DFT) calculations \cite{wang1,jang1}, assuming that the graphene monolayer is strictly aligned with the Ru lattice, have concluded that the ripples are mostly structural in origin, predicting a geometrical corrugation of 1.5 {\AA}. On the contrary, a very recent \textit{ab initio} calculation that assumes a slightly different registry between graphene and Ru(0001), gives a corrugation value of 0.24 {\AA} \cite{Peng}. A first fit to surface X-ray Diffraction data suggest a corrugation of 1.5 {\AA} \cite{martoccia1}, using the same technique (XRD) a recent fit \cite{martoccia2} gives a corrugation of 0.82 {\AA}. Finally in a recent work using Low Energy Electron Diffraction (LEED) a corrugation value of 1.5 {\AA} \cite{moritz1} have been proposed. It is worth to mention the limited sensitivity of diffraction techniques to the position of the light Carbon surface atoms as compared with the Ruthenium ones. Recently, a model calculation \cite{sutter2} has found, contrary to the previously mentioned DFT calculations \cite{wang1,jang1}, that the structural corrugation of the moir\'{e} pattern  is comparable to the atomic corrugation of the carbon atoms. Considering its paradigmatic nature as a prototype nanostructured graphene system, it is rather upsetting that there is still a lack of consensus regarding the structure of gr/Ru(0001).

In this work we study the geometric and electronic structure of graphene epitaxially grown under UHV conditions on Ru(0001) by low-temperature STM and Helium Atom Scattering (HAS). The high-resolution STM images of large, atomically perfect domains  allow us to determine that the graphene lattice is rotated by 0.5$^{\circ}$ with respect to the Ru lattice. The moir\'{e} pattern, which is rotated by 5.0$^{\circ}$ with respect to the Ru lattice in the STM images, acts as a magnifying lens for the small angle of rotation between the two atomic lattices \cite{diaye2}. The intensity of the He diffraction peaks, allows us to quantify accurately the apparent corrugation of the total charge at the external surface (at about the same distance explored by STM), i.e. the closest experimental measure of the geometric corrugation of the ripples, which turns out to be of the substantially smaller than predicted \cite{wang1, jang1} or reported by XRD \cite{martoccia1, martoccia2} and LEED \cite{moritz1}. The much larger apparent STM corrugation depends strongly on the bias voltage, and can even be inverted above +2.6 V, revealing that, in addition to the geometric corrugation, a much stronger electronic corrugation exists.

\section{Results and discussion}

The STM and HAS experiments were carried out in different UHV chambers with base pressures of 1$\times$10$^{-10}$ mbar, equipped with standard facilities for metal surface preparation, ion gun and mass spectrometer, gas exposure, Low Energy Electron Diffraction and Auger Electron Spectroscopy. Atomically clean, bulk C depleted, crystalline Ru(0001) surfaces were prepared by standard sputter/anneal procedures followed by oxygen exposure at 1150 K and a final flash to 1400 K in UHV, which resulted in large terraces displaying atomic resolution, separated by monoatomic steps.

In order to minimize the possible influence of defects and domain boundaries in the observed registry between the graphene overlayer and Ru(0001) that may induce distortion on the moir\'{e} superstructure \cite{marchini1}, special care was taken to clean the bulk of the Ru crystal from C by repeatedly exposing it to 3$\times$10$^{-7}$ mbar of O$_2$ at 1150 K and flashing it  to 1400 K. Different partial pressures of ethylene were explored during the growth. The samples of graphene were prepared on multiple occasions on two different Ru(0001) single crystals following the procedure that yields almost perfect overlayers and larger domains: the Ru crystals, kept at 1150 K in Ultra High vacuum (UHV), were exposed to ethylene at pressures of 2$\times$10$^{-7}$ mbar for 3 minutes (48 L, 1 L=1.33$\times$10$^{-6}$ mbar s) or 3$\times$10$^{-8}$ mbar for 10 minutes (24 L). These exposures were enough to saturate the surface. The temperature was held at 1150 K for further 2 minutes after removing the C$_2$H$_4$ gas from the chamber.

These optimum conditions were determined from the combined STM and HAS analysis of the grown graphene films, which display almost atomically perfect domains covering completely the surface. Figure 1(a) shows an STM image of a single domain with a lateral size of 2000 {\AA}. The periodic bright dots are the ripples of the moir\'{e} pattern of epitaxial gr/Ru(0001). Only one domain is visible in the image as reflected in the sharpness and order of the spots in the 2D Fast Fourier Transform (FFT) of the image shown on the inset. Note that the ripples in all the terraces are aligned exactly along the same direction. The only defects present in the image are bubbles of Argon (green circle) buried in the substrate during the ion bombardment of the Ru(0001) crystal \cite{gsell,calleja1} that appear as cloudy regions. They do not perturb significantly the periodicity of the moir\'{e} superstructure.

In larger images, such as the one shown in Figure 1(b), the presence of dislocations in the moir\'{e} structure (black circles) is a clear indication of the existence of several domains. In order to elucidate if the domains are due either to the coalescence of graphene islands with a different registry respect to the Ru(0001) lattice or to misalignments between the graphene lattices on the different domains, we have calculated the FFT of the image \cite{wsxm}. The result is shown in the inset in Figure 1(b), there is only one set of spots in an hexagonal arrangement corresponding to the moir\'{e} superstructure. From the comparison with the FFT shown in Figure 1(a) it is clear that these spots are elongated. The origin of this deformation is the presence of different domains with a small misalignment between them. The angular spread of the moir\'{e} patterns can be determined by measuring the angle between the green lines that connect the (0,0) point with the boundaries of the elongated (1,1) spot and turns out to be 10$^{\circ}$.

\begin{figure}
\includegraphics*[width=168mm]{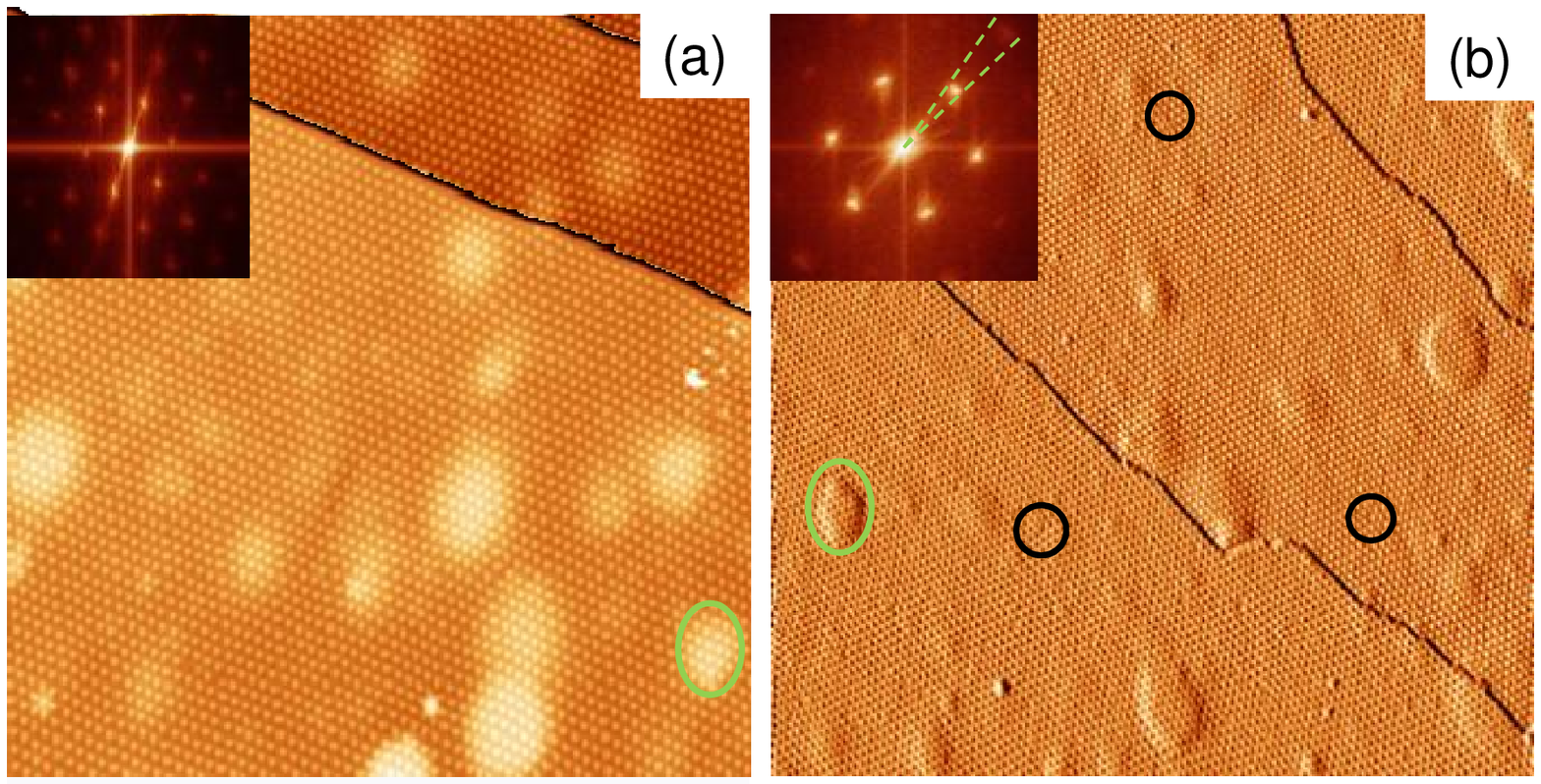}
\caption{(a) 2000$\times$2000 {\AA}$^2$ STM image of graphene on Ru(0001). The color scale has been adjusted independently on every terrace. The ordered white dots correspond to the maxima of the moir\'{e} superstructure. The green circle encloses a subsurface Argon bubble in the Ru(0001) substrate. The inset shows the Fourier transform of the image showing the existence of a single hexagon and well defined spots, a clear indication of long range order on the moir\'{e} superstructure. (b) 3000$\times$3000 {\AA}$^2$ STM image of graphene on Ru(0001). The STM images have been differentiated along the horizontal direction. Black circles mark some of the dislocations present on the moir\'{e} superstructure. The inset shows the 2D Fast Fourier Transform of the image containing the first hexagon of the moir\'{e} superstructure. The spot corresponding to the moir\'{e} periodicity is elongated showing an angular distribution of $\pm$5$^{\circ}$ marked by green lines.}
\label{figure1}
\end{figure}

Figure 2 (a) shows a high resolution STM image of the graphene monolayer, with its characteristic triangular array of bumps separated (29.3 $\pm$ 0.8) {\AA} and the simultaneously resolved atomic C lattice. The green dotted line indicates the high-symmetry [11\={2}0] direction of the carbon lattice and the blue dotted line the corresponding high symmetry direction of the hexagonal moir\'{e} pattern. The atomic rows are clearly not aligned with the ripples in all the areas free from defects that have been examined. The angle between both directions is $\varphi$$_{gr,moire}$ = 4.5$^{\circ}$ $\pm$ 0.5$^{\circ}$. Similar misalignment between the carbon lattice and the ripples can be observed in LEED patterns measured with a Low Energy Electron Microscope (LEEM) on a single domain island of epitaxial gr/Ru(0001) but went unnoticed \cite{sutter1, loginova1}. For diffraction patterns measured with conventional LEED on continuous graphene films the co-existence of the misaligned domains makes the diffraction spots wider but centered on the high symmetry directions (see insets in Figure 1(b) and in Figure 4(a)) rendering difficult to identify such a small rotation between the atomic lattice and the moir\'e superstructure. The characterization done in real space by means of atomically resolved STM images published so far show only one or two unit cells of the moir\'e superstructure, making difficult to identify such small rotation but in the high quality ones \cite{gao1, pan1} it is easy to see the rotation between the atomic rows and the ripples. This fact, however, has not been noticed.

In Figure 2 (b) the 2D Fast Fourier Transform of the topographic image shows simultaneously the periodicity of the moir\'e superstructure and the carbon lattice (green circles). From the power spectrum it is clear the co-existence of  two different periodicities rotated respect each other and with a distance ratio of 1/12 approximately. Figure 2 (c) shows the profile in the spectral density drawn in the FFT image along the line connecting the spots corresponding to the graphene atomic periodicity i.e. along the high symmetry direction of the substrate (green circles in panel (b)). The peaks corresponding to the atomic periodicity can be seen at both ends of the spectral density (marked with red arrows). Due to the small rotation between the carbon atomic structure and the moir\'{e} superstructure, only the first two peaks corresponding to the ripples periodicity can be seen. On the contrary a line profile in the spectral density rotated by 5$^{\circ}$ pass through all the peaks corresponding to the periodicity of the moir\'{e} superstructure as shown in Figure 2(d) (blue profile).

\begin{figure}
\includegraphics*[width=168mm]{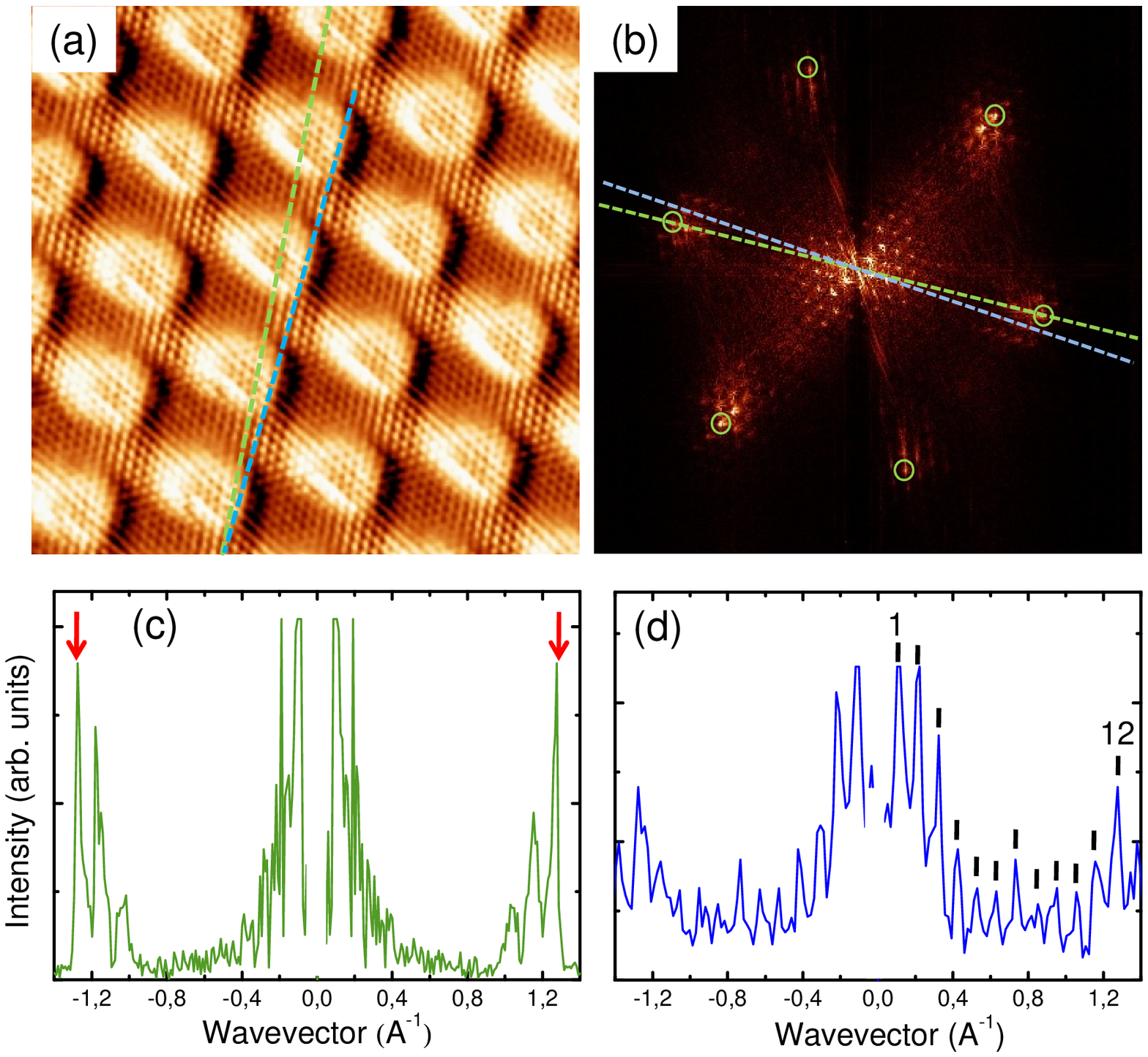}
\caption{(a) Atomically resolved STM image (130$\times$130 {\AA}$^2$) of the complete graphene monolayer grown on Ru(0001) recorded at 4.6 K in an area free of defects (image setpoint V$_s$ = +1mV and I$_t$ = 1 nA). The green dotted line in the topographic image follows the high-symmetry [11\={2}0] direction of the C atomic rows in the graphene layer. The blue dotted line shows the direction of the moir\'{e} superstructure. (b) Power spectrum of the STM image shown in (a). The green circles mark the spots due to the graphene atomic periodicity; (c) Spectral density along the high symmetry direction of the atomic structure marked with a green line on panel (b). (d) Spectral density along the high symmetry direction of the moir\'{e} pattern marked with a blue line on panel (b).}
\label{figure2}
\end{figure}

The rotation between the C and the moir\'{e} lattices for gr/Ru(0001) reflects in an amplified fashion the misalignment between the graphene monolayer and the underlying Ru lattice \cite{diaye2}. In order to determine directly the epitaxial relationship of the graphene lattice with the Ru substrate, the growth of graphene has been stopped just below the completion of the monolayer \cite{borca1} allowing the existence of small patches of the Ru substrate not yet covered by graphene. Since it has proven very difficult to obtain simultaneously atomic resolution in both the graphene islands and the Ru substrate, the surface has been exposed to a small O$_2$ dose. Oxygen does not adsorb at 300 K on graphene \cite{borca1}, but it does on the clean Ru patches, forming a well known (2$\times$2) superstructure epitaxially aligned with the Ru lattice, which is imaged with large corrugation by STM \cite{calleja2}. The O(2$\times$2) superstructure is, thus, employed as a ruler to reveal the epitaxial relationship between graphene and Ru.

Figure 3 shows an STM image of the edge of a graphene island with its characteristic bumps due to the moir\'{e} superstructure and the simultaneously resolved (2$\times$2) arrangement of oxygen atoms adsorbed on the adjacent Ru area. The green line follows one of the high-symmetry directions of the (2$\times$2) superstructure, i.e. a symmetry direction of the Ru lattice. It deviates from the line formed by the bumps by 5.0$^{\circ}$ $\pm$ 0.5$^{\circ}$. The same happens for the equivalent high symmetry directions. This is confirmed by the Fourier transform of the STM image shown in the inset of Figure 3. The broad spots corresponding to the (2$\times$2) superstructure are visible, while the sharper ones reflect the graphene superstructure. The moir\'{e} superstructure is rotated with respect to the O(2$\times$2) pattern and, accordingly, with respect to the Ru lattice by $\varphi$$_{Ru,moire}$ = 5.0$^{\circ}$ $\pm$ 0.5$^{\circ}$.

Due to the magnifying effect of the moir\'{e} pattern \cite{diaye2}, that is 9.96 for gr/Ru(0001) system, the misalignment of the C and Ru lattices can be determined with high precision from the observed angles between the moir\'{e} superstructure and the Ru or C lattices and it turns out to be $\varphi$$_{gr,Ru}$ = 0.5$^{\circ}$ $\pm$ 0.05$^{\circ}$. This small rotation might explain the contradiction between the periodicity (25$\times$25)\cite{martoccia1} obtained with XRD measuring the deformation of the last Ruthenium layers and the moir\'e periodicity obtained from the STM images \cite{marchini1}. The superposition of the graphene and Ru(0001) lattices without any distortion and with the [\={1}010] directions rotated by 0.5° produces a (24$\times$24) periodicity when the registry with the Ru(0001) is taken into account. This larger periodicity probably reduces the compression needed to accommodate the graphene overlayer over Ru(0001) ($\sim$0.78\% from DFT calculations \cite{jang1}) and, presumably, also the structural crumpling of the graphene overlayer.

\begin{figure}
\includegraphics*[width=84mm]{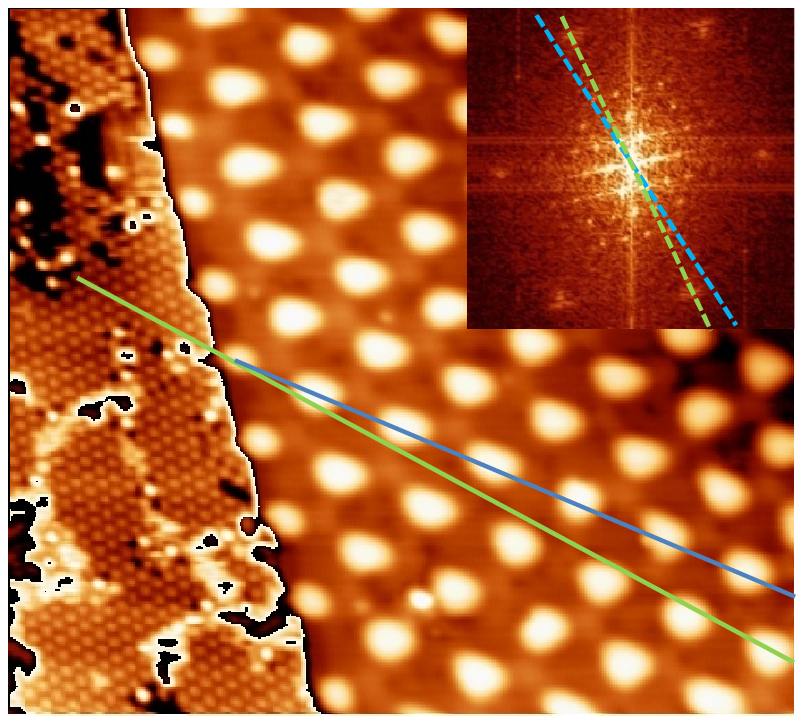}
\caption{300$\times$270 {\AA}$^2$ STM image of the edge of a large graphene island grown on Ru(0001) coexisting with a (2$\times$2) superstructure of oxygen chemisorbed on the Ru patches not covered by graphene (image setpoint Vs = -1V and It = 0.1 nA). The green line guides the eye along the high symmetry direction of the O(2$\times$2) superstructure. The blue line guides the eye along the moir\'{e} pattern. The color scale has been adjusted independently on the (2$\times$2) superstructure and on the graphene island. Inset: Fourier transform of the STM image.}
\label{figure3}
\end{figure}

The structural corrugation of the graphene monolayer can be determined by HAS. The advantages of using a beam of He atoms of thermal energy (10-100 meV) as a probe of the surface structure are its combination of low energy with short wavelength, its inert and neutral character and its large cross-section for defects. As a consequence, HAS is a unique non-destructive and surface sensitive technique, with high sensitivity to low mass atoms, such as C, or light adsorbates, such as H \cite{farias1}. The actual diffraction grating is the periodic modulation of the repulsive part of the He-graphene potential at the energy of the incoming He atoms. These classical turning points define a corrugation function $\xi$(x,y), which is a replica of the total surface electron density profile at about 2-3 {\AA} above the nuclei. The amplitude of the corrugation function dictates the intensity of the diffraction beams. The general problem of calculating diffraction intensities for a given scattering geometry and corrugation function consists in solving the time-independent Schr\"odinger equation with a realistic soft potential V({\bf r}). This problem can be solved exactly in the most general case using the close-coupling method \cite{wolken1}. In our case, we have solved the close-coupling equations applying the procedure developed by Manolopoulos et al. \cite{manolopoulos}, which achieves convergence much faster than the method originally proposed by Wolken \cite{wolken1} and is therefore more appropriate for calculations of large unit cells like the moir\'e of graphene on Ru(0001).

\begin{figure}
\includegraphics*[width=168mm]{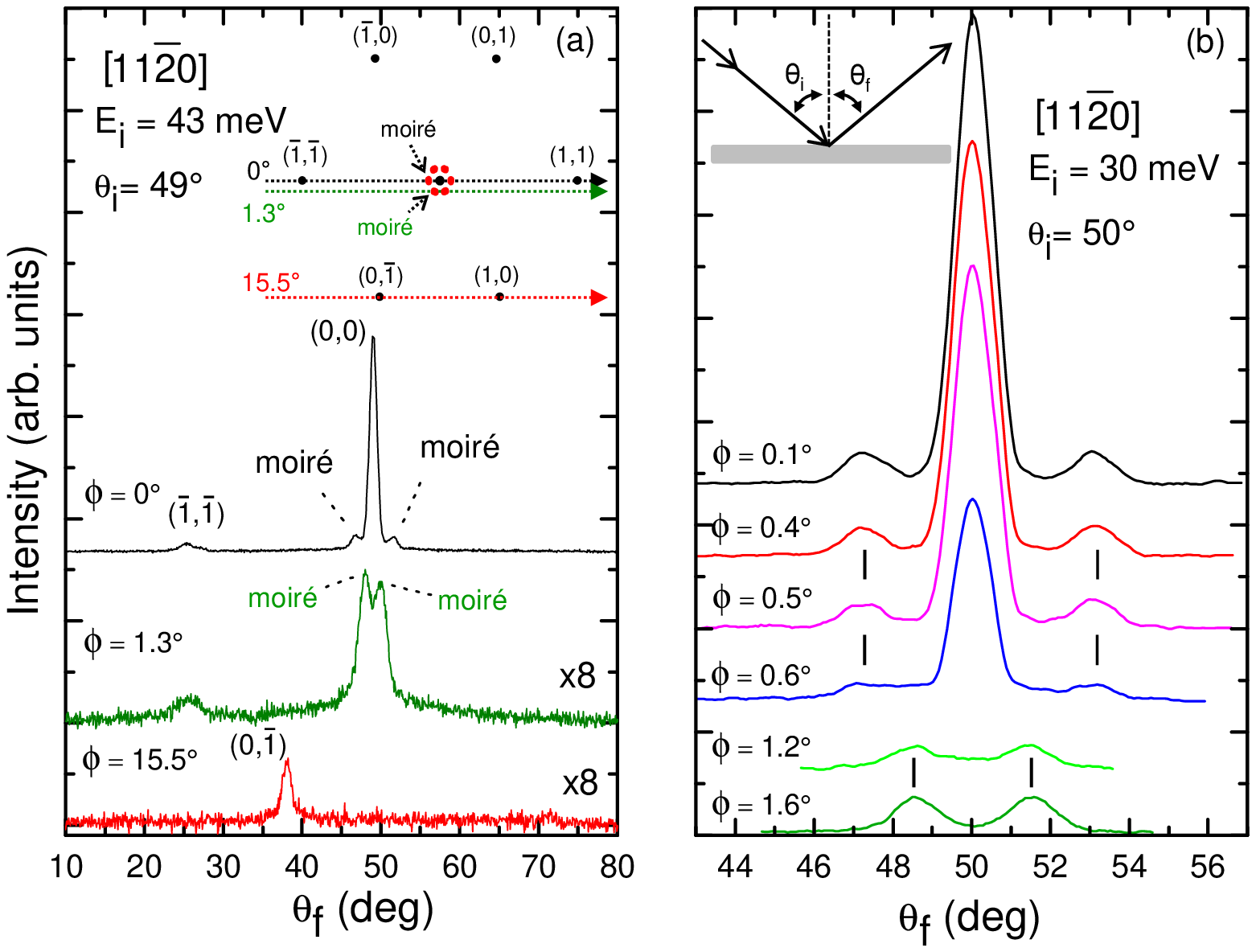}
\caption{(a) Angular distribution of the He atoms scattered off graphene/Ru(0001). The azimuthal angle, $\phi$, is defined respect to the [11\={2}0] direction. The energy of the incident He atoms is 43 meV, the incident angle respect to the surface normal is 49$^{\circ}$ and the surface is kept at 100K during the scattering experiments. The inset indicates the two rotated ($\pm$5$^{\circ}$) moir\'e superstructure reciprocal lattices (red hexagons) and the graphene atomic lattice (black hexagon). (b) Evolution of the intensity of the specular beam and the first diffraction peaks of the moir\'e superstructure as a function of the azimuthal angle $\phi$. As we move out-of-plane, $\phi$ increases, and the intensity of the specular beam and the first order diffraction peaks decreases. For $\phi$ larger than 1.2$^{\circ}$, the specular peak disappears, and the first out-of-plane diffraction peaks from the moir\'e superstructure appear. The inset shows the in-plane scattering geometry.}
\label{figure4}
\end{figure}

Figure 4(a) shows the in-plane ($\phi$=0$^{\circ}$) (black line) and out-of-plane ($\phi$=1.3$^{\circ}$, 15.5$^{\circ}$) (green and red lines respectively) He-diffraction spectra from a graphene overlayer grown on Ru(0001) measured along the [11\={2}0] direction, with a He beam energy of 43 meV. The intensity of the specularly reflected (0,0) peak depends on the structural perfection of graphene, reaching 25\% of the incident beam for highly perfect layers. The in-plane He-diffraction spectrum shows both the first peaks of the moir\'{e} superstructure (close to the specular beam) and the (\={1},\={1}) of the C lattice. It is worth mentioning that the appearance of the first moir\'{e} peaks and the (\={1},\={1}) in the same in-plane scan cannot be taken as an indication that the corresponding lattices are aligned. The inset in Figure 4(a) shows the reciprocal lattices of the graphene atomic structure (black dots) and the moir\'e superstructure (red dots). For the moir\'e superstructure two reciprocal lattices, corresponding to two domains rotated $\pm$5$^{\circ}$ are shown. With the existing angular resolution (1.4$^{\circ}$), estimated from the full width at half maximum (FWHM) of the measured peaks, and the small deviation between the atomic rows and the moir\'{e} pattern a scan through the atomic lattice peaks in the reciprocal space unavoidably passes through the first two peaks of the moir\'{e} superstructure (note the inset in Figure 1(b)).

\begin{figure}
\includegraphics*[width=168mm]{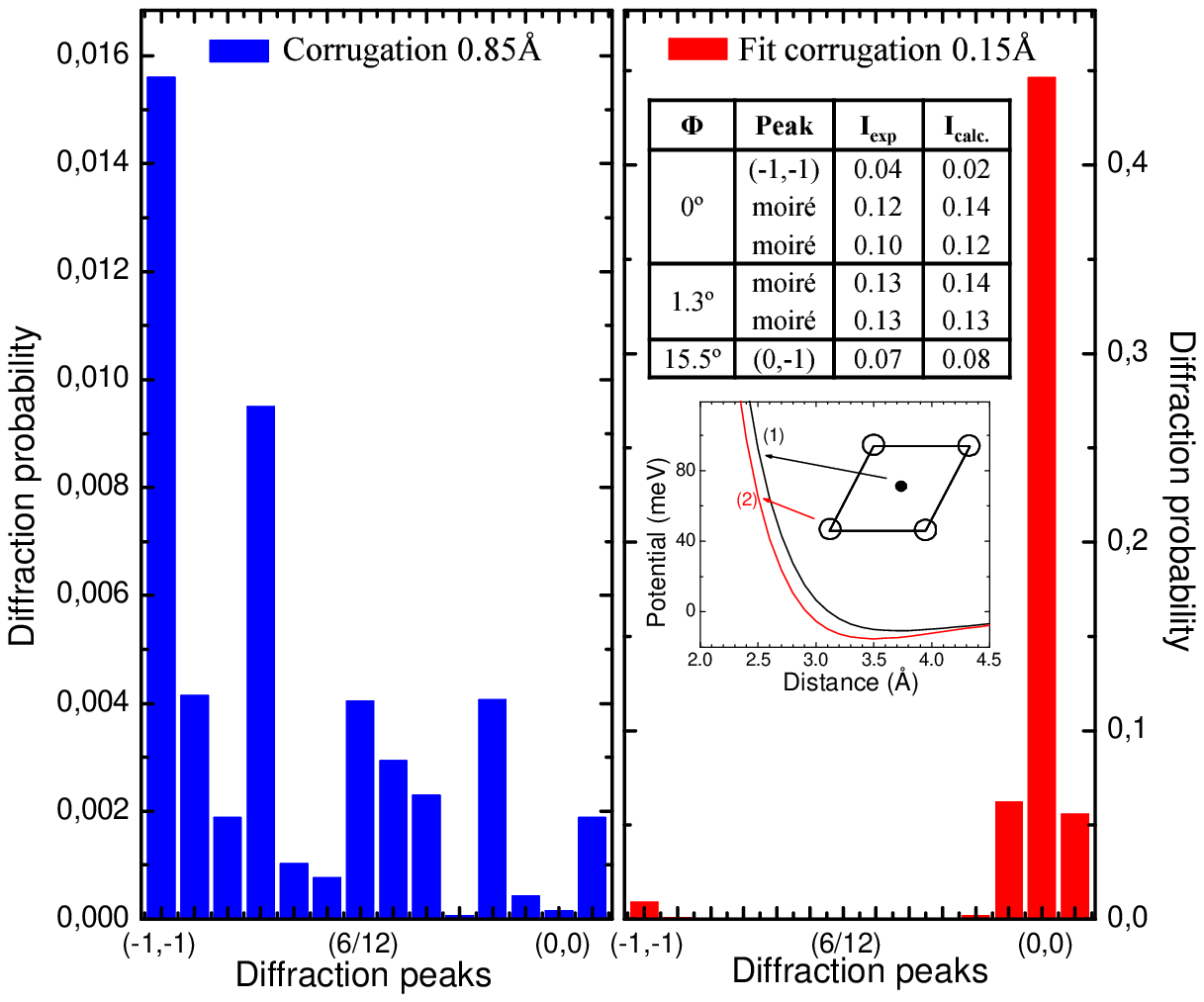}
\caption{Close-coupling calculations of the in-plane diffraction probabilities for two different corrugation amplitudes of the  moir\'e superstructure, 0.85 {\AA} (left panel) and 0.15 {\AA} (right panel) with an incident energy of 43 meV. Note the large differences in the relative values of diffraction probabilities with respect to the specular beam (0,0). The table in the right panel compares the best fit intensities obtained with close-coupling calculations with the experimental data of the moir\'e superstructure shown in Figure 4(a), corresponding to $\theta _i$=49$^{\circ}$ and E$_i$= 43 meV. The intensities are normalized to the specular peak. Inset: real space potential profiles at the corner (1) and in the middle (2) of the moire unit cell.}
\label{figure5}
\end{figure}

The relative intensity of the different diffraction peaks with respect to the specular peak determines unequivocally the corrugation function, $\xi$(x,y), i.e. the corrugation of the constant charge density contour where the He atoms are reflected. The potential V({\bf r}) has been modeled by a discrete Fourier series,
\begin{equation}
         V({\bf r}) = \sum_{\bf G}  A_{\bf G} e^{-{B_{\bf G}}z} e^{i {\bf G} {\bf R} (x,y)},
\end{equation}
where {\bf R} is the component of {\bf r} in the surface plane and {\bf G} a reciprocal lattice vector. $A_{\bf G}$ and $B_{\bf G}$ are the coefficients corresponding to the amplitude and exponential attenuation, respectively, of the different terms used in the fitting procedure (two terms in our case). $V_{{\bf G}=0}$ denotes the laterally averaged potential, and $V_{{\bf G}\neq 0}$ the Fourier coefficients of the periodic part of the potential. $V_{{\bf G}=0}$ has been modelled by a two parameter Morse potential,
\begin{equation}\label{morse}
         V(z) = D(e^{-2{\alpha}(z-z_0)} - 2e^{-{\alpha}(z-z_0)}),
\end{equation}
using for D (potential well depth) and $\alpha$  (range parameter) the values D = 14 meV and $\alpha$ = 1.15 {\AA}$^{-1}$, which were derived from the selective adsorption resonances reported for the He-graphite(0001) interaction by Boato et al. \cite{boato}. The unit cell was modelled including two different coefficients
$A_{\bf G}$, corresponding to the graphene atomic periodicity and the moir\'e periodicity (but not the small misalignment between them), whereas the same coefficient  $B_{\bf G}$ was used for both, and the corrugation function $\xi$(x,y) was determined by fitting the measured in-plane and out-of-plane diffraction intensities by means of a trial and error procedure. In the table show as inset in Figure 5 (right panel) we compare the experimental intensity of the diffraction peaks, normalized to the specular intensity, with our best fit. The corresponding corrugation function obtained for the moir\'e superstructure periodicity has a maximum corrugation amplitude of 0.15 {\AA}, one order of magnitude smaller than the value theoretically predicted \cite{wang1,jang1} for the aligned graphene overlayer. This low value was expected from the similar intensities observed in Figure 4(a) for the C-associated (\={1},\={1}) He diffraction peak and the ones coming from the moir\'{e} superstructure. A possible explanation for the high corrugation predicted by the DFT calculations is the forced alignment of the graphene lattice to the underlying Ru lattice imposed in the structural model used in the DFT calculations \cite{wang1,jang1}, not allowing the system to explore all the possible mechanisms to relax the atomic positions and producing an artificially large structural corrugation. The energy dependence of the corrugation amplitude can be found in the inset of Figure 5, which shows the potential in real space, giving a surface corrugation of 0.15 {\AA} at $\rm E_{iz} =20$ meV.From this graph one can also estimate that the corrugation amplitude remains almost unchanged around 0.15 {\AA} in the range of incident energies between 30-100meV.

The previous fit has only sense if the total intensity of diffracted and specular peaks is proportional to the scan performed with the existing resolution. This could be wrong if the width of the diffracted peaks would be substantially larger than the width of the specular peak in the azimuthal direction. In order to establish that the relative intensities of moir\'e/specular peaks measured in a line scan reflect the true relationship between intensities, the width of the peaks has been determined by performing additional out-of-plane measurements. Figure 4(b) shows the evolution of the intensity of the specular beam and the first order diffraction peaks corresponding to the moir\'e superstructure when the He detector is moved in tiny steps in the out-of-plane direction. When the detector is moved out-of-plane, both the specular beam and the in-plane diffraction peaks corresponding to the moir\'e periodicity get smaller in such a way that the intensity ratio between the specular peak and the first order diffraction ones remains almost constant (curves measured with $\phi$ = 0.1$^{\circ}$, 0.4$^{\circ}$, 0.5$^{\circ}$ and 0.6$^{\circ}$ in Figure 4(b)). For larger $\phi$ values (1.2$^{\circ}$ and 1.6$^{\circ}$), we observe the appearance of the first order, out-of-plane diffraction peaks corresponding to the moir\'e superstructure. We explored in this way  $\phi$ values in the range $\pm$20$^{\circ}$, which correspond in the reciprocal space to more than one unit cell of the atomic periodicity, since the first order out-of-plane peaks corresponding to the atomic periodicity appear at 15.5$^{\circ}$, as can be seen in Figure 4(a) (red curve). These measurements allow us to conclude that the FHWM of diffraction peaks corresponding to the (1$\times$1) and the moir\'e superstructure are comparable. This means that there is no more intensity than the one we show in these spectra and, therefore, our comparison in intensity between the specular peak and the diffraction ones is meaningful.

\begin{figure}
\includegraphics*[width=120mm]{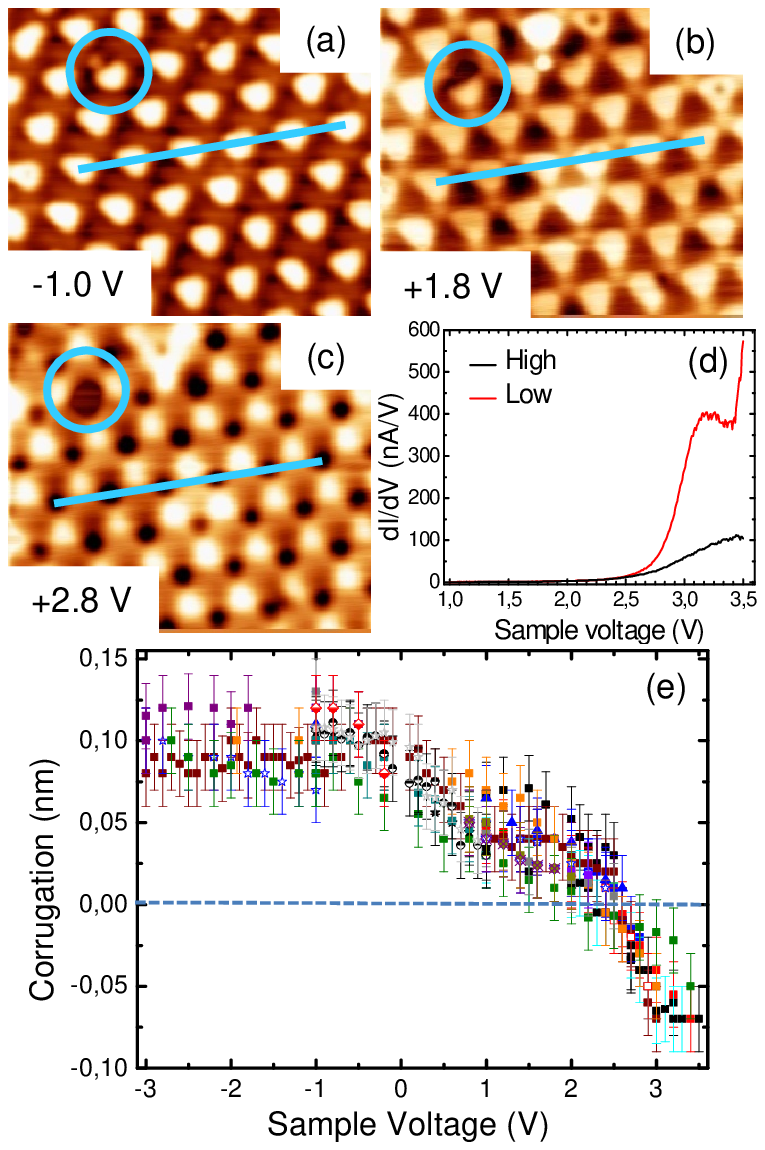}
\caption{Inversion of the contrast in STM images of graphene/Ru(0001). (a), (b) and (c) show topographic STM images at three selected sample bias voltages. (d) Local Tunnel Spectroscopy recorded with the tip located on top of the high (black curve) and low (red curve) areas of the moir\'{e} pattern in panel (a). (e) Voltage dependence of the apparent topographic corrugation for many different experimental conditions (tip, tunneling current, samples and temperature). The data clearly show a contrast inversion for bias voltages larger than +2.6 eV.}
\label{figure6}
\end{figure}

Finally, we have studied the sensitivity of our best-fit results to the Morse parameters. Briefly, we performed several fits varying D between 12-18 meV, and $\alpha$  from 0.8 to 1.4 $\rm {\AA}^{-1}$.  As a result, we found that our data could always be fitted with a corrugation amplitude in the range 0.15-0.20 $\rm{\AA}$. This means that with our current data, although we cannot refine further the best-fit parameters, we can safely state an upper limit to the maximum corrugation amplitude of this surface at 0.20 $\rm{\AA}$. Further support to this conclusion comes from calculations performed for the best-fit D and $\alpha$  values, which show that the intensity of the first order moir\'{e} peak strongly depends on the corrugation's value, reaching $\sim 50 \%$ of the specular peak for a corrugation amplitude of just 0.20 $\rm{\AA}$. In addition, for this value already the second order moir\'{e} peak starts to be visible. Since we don't see this peak in the experiment, we can conclude that the corresponding corrugation amplitude cannot be larger than 0.20~$\rm{\AA}$.

We have also studied the energy dependence of the diffraction intensities by performing measurements at higher incident energies (up to 125 meV)  which were complemented by close-coupling calculations. These measurements  show that the relative intensity of the first moir\'{e} peak (compared to the specular one) increases from ca. 12\% at 30 meV to ca. 27\% at 125 meV, while the intensity of the (\={1},\={1}) peak remains almost unchanged in the same  energy range. This observation already shows that the energy dependence of the corrugation amplitude, if any, can be only very weak. We have also quantified this by fitting the data at the highest energy (varying also slightly the parameters of the Morse potential around our best-fit values at $\rm E_i=43$ meV), obtaining a corrugation amplitude of 0.17~$\rm{\AA}$, i.e. very similar to the value we got at lower energies.

In order to check how sensitive are our close-coupling calculations to changes in the surface corrugation, we performed calculations using two different values for the corrugation amplitude of the moir\'e superstructure,  without changing any other parameter in our model. In Figure 5 (right panel) we show the result of such a calculation for a moir\'e corrugation of 0.15 {\AA}. Most of the intensity (up to 44$\%$) is concentrated in the specular beam, and only around 6$\%$ goes to the first in-plane diffraction peaks due to the moir\'e superstructure. This calculated diffraction spectrum closely reproduce our in-plane measurements, as can be seen from the comparison of the normalized intensities in the table shown in Figure 5 (right panel). As we increase the corrugation of the moir\'e superstructure more intensity is diffracted away from the specular beam \cite{farias1}. If the corrugation of the moir\'e superstructure is increased up to 0.82 {\AA} \cite{martoccia2}, the new calculated diffraction spectra (left panel in Figure 5) show a very weak specular peak with less than 0.1 $\%$ of the total diffraction intensity,  and most of the intensity is distributed among the many diffraction peaks due to the periodicity of the moir\'e superstructure. This new calculated diffraction spectra does not reproduce any of the main features of our experimental data, in which the most intense peak is the specular one and the diffraction peaks due to the moir\'e superstructure rather weak (see Figure 4(a)(b)).

It should be mentioned that on some metallic substrates \cite{rieder1,rieder2,petersen1} the He diffraction technique underestimates the surface corrugation due to the interaction of the He 1s electrons with the metallic surface electronic structure producing the so called "anticorrugation effect". Calculations suggest an upper limit for this "anticorrugation effect" of the order of 0.2 {\AA} \cite{petersen1, annett1} for metallic surfaces. Without a detailed calculation we cannot exclude the existence of this phenomenon in this system, where the electronic structure is strongly modulated \cite{amadeo1}. Therefore our estimate might be a lower bound to the actual corrugation and, we can safely conclude that the structural corrugation of the moir\'{e} superstructure is in between 0.15 {\AA} and 0.4 {\AA}, still much smaller than the value predicted by the DFT calculations \cite{wang1,jang1} or the experimental values determined by X-ray surface diffraction \cite{martoccia1} or low energy electron diffraction \cite{moritz1}. A possible problem in these diffraction techniques is the low sensitivity to light atoms (i.e. carbon) compared to the transition metal substrate (ruthenium) which makes difficult the precise location of the C atoms, specially in view of the size and complexity of the unit cell with more than 1000 atoms for the periodicity suggested by X-ray surface diffraction \cite{martoccia1}.

The STM images measured on this surface present always the same lateral periodicity but depending on the bias voltage the apparent corrugation of the moir\'e superstructure changes strongly with voltage and can even be inverted \cite{pan1,borca3}. If the corrugation measured with the STM has its origin in a modulated electronic structure \cite{amadeo1,amadeo2} in addition to a physically corrugated graphene layer, the value measured with the STM should change dramatically with the sample bias voltage. Figure 6 (a),(b),(c) show STM topographic images recorded at representative sample biases and at 4.6 K. A defect in the superstructure encircled in blue provides with an absolute reference against possible drift between the images. The blue line highlights that the regions visualized as bumps at negative sample voltages (occupied states), are seen as depressions at +2.8 V (empty states). Panel (e) shows the voltage dependence of the apparent topographic corrugation for many different experimental conditions (tip, tunneling current, samples and temperature). When imaging occupied states, the apparent corrugation is rather constant ($\sim$1 {\AA}), as expected because electrons at the Fermi level contribute the most to the tunneling current, but the corrugation decreases continuously when injecting electrons in the empty states of graphene and becomes negative above +2.6 V. This behavior is fully reversible and do not depend on the sample temperature in the range between 4.6K and 300K.

Spatially resolved scanning tunneling spectroscopy (Figure 6 (d)) shows that the inversion of the contrast is due to the appearance of an intense peak in the empty states at +3.0 eV that is localized in the apparent depressions of the topographic images recorded at negative bias voltage. As demonstrated elsewhere, the origin of this state is the hybridization of the graphene first empty state with the Ru conduction band \cite{borca2}.

\section{Conclusions}

In conclusion, we have found for gr/Ru(0001) that the graphene lattice is rotated with respect to the Ru lattice, a fact overlooked in previous theoretical calculations. The structural corrugation of the total charge of the ripples as determined by HAS is 0.15 {\AA}, substantially smaller than predicted by DFT calculations. Most of the apparent corrugation in STM images is of electronic origin, being this component so strong that it leads to a reversible inversion of the contrast above +2.6 V. This is originated by an empty electronic state derived from the Ru bands and spatially localized in the valleys of the structural corrugation. The electronically corrugated graphene overlayer is a self-nanostructured playground where new physics and spatially organized chemistry is bound to appear.

\section{Acknowledgments}

Financial support by the Ministerio de Educaci\'on y Ciencia through projects CONSOLIDER-INGENIO 2010 on Molecular Nanoscience and FIS2007-61114 and Comunidad de Madrid through the program NANOBIOMAGNET S2009/MAT1726 is gratefully acknowledged. A.P. thanks the Ministerio de Educaci\'on for financial support.

\end{document}